\documentclass[aps, prd, onecolumn, nofootinbib, preprint]{revtex4-2}

\usepackage[colorlinks=true, linkcolor=blue, citecolor=blue, urlcolor=blue]{hyperref}
\usepackage{amsmath, graphicx, braket, orcidlink, subfigure}

\renewcommand{\Im}{\operatorname{Im}}
\DeclareMathOperator{\Tr}{\mathrm{Tr}}

\begin{document}

\title{
Baryogenesis in $SU(2)_{L}$ multiplet models
}

\author{Kiyoto Ogawa\,\orcidlink{0009-0001-7115-7107}}
\email{ogawa.kiyoto.f8@s.mail.nagoya-u.ac.jp}
\affiliation{Department of Physics, Nagoya University, Furo-cho Chikusa-ku, Nagoya 464-8602 Japan}

\author{Masanori Tanaka\,\orcidlink{0000-0002-1303-7043}}
\email{tanaka@pku.edu.cn}
\affiliation{Center for High Energy Physics, Peking University, Beijing 100871, China}

\begin{abstract}

We investigate baryogenesis in Standard Model (SM) extensions with new $SU(2)_L$ multiplet fields.
We focus on sphalerogenesis, in which the baryon asymmetry of the Universe (BAU) is generated through the gradual decoupling of CP-violating electroweak sphaleron-like processes.
We show that the observed BAU can be reproduced when the new fields possess CP-violating Yukawa interactions, which leave a CP-violating dimension-six operator involving the $SU(2)_L$ gauge fields at low energies.
As representative examples, we study models with fermionic $SU(2)_L$ quintuplets and septuplets, and find that these field masses should be $\mathcal{O}(1)\,\mathrm{TeV}$ to explain the BAU. 
We also show that viable parameter regions for the BAU are consistent with current bounds on the electron electric dipole moment and thoroughly probed by future measurements such as ACME III and by mono-lepton searches at the HL-LHC.
Our results provide a concrete and phenomenologically testable ultraviolet completion of sphalerogenesis.

\end{abstract}

\maketitle 
\newpage

\tableofcontents

\section{Introduction}

The origin of the baryon asymmetry of the universe (BAU) remains one of the central questions in particle cosmology. 
Measurements of the cosmic microwave background strongly constrain the baryon-to-entropy ratio as~\cite{Planck:2018vyg}
\begin{align}
\label{eq:BAU}
8.41 \times 10^{-11} < \frac{n_{B}}{s} < 8.75 \times 10^{-11} \,. 
\end{align}
If the universe started from a baryon-symmetric initial state, a mechanism satisfying the Sakharov conditions is required to reproduce the observed baryon asymmetry in Eq.~\eqref{eq:BAU}~\cite{Sakharov:1967dj}. 
The Sakharov conditions consist of three requirements: (i) baryon number violation, (ii) C and CP violation, and (iii) departure from thermal equilibrium.

To satisfy the third Sakharov condition, the dynamics of the electroweak (EW) phase transition is often invoked, as in EW baryogenesis~\cite{Kuzmin:1985mm}. 
However, lattice simulations have shown that the EW phase transition in the Standard Model (SM) is a smooth crossover for the observed Higgs boson mass~\cite{Kajantie:1996mn,Csikor:1998eu,Aoki:1999fi,DOnofrio:2014rug,DOnofrio:2015gop}. 
Therefore, the SM cannot realize a first-order EW phase transition, motivating physics beyond the SM to provide the required departure from thermal equilibrium.

On the other hand, a new possibility for baryogenesis during smooth EW symmetry breaking was proposed~\cite{Kharzeev:2019rsy,Hong:2023zrf,Tanaka:2025cpw}. 
In this scenario, the Sakharov conditions are satisfied solely through EW sphaleron dynamics: (i) baryon number violation by EW sphaleron processes, (ii) CP asymmetry in EW sphaleron transitions, and (iii) departure from equilibrium through the gradual decoupling of sphaleron-like processes. 
This mechanism is called sphalerogenesis~\cite{Tanaka:2025cpw}. 
In Ref.~\cite{Kharzeev:2019rsy}, the CP asymmetry in the sphaleron process was estimated from interactions between the EW gauge fields in the sphaleron configuration and fermions emitted via sphaleron explosions. 
Using this result, it was found that sphalerogenesis within the SM produces a baryon asymmetry about three orders of magnitude smaller than the observed value in Eq.~\eqref{eq:BAU}~\cite{Kharzeev:2019rsy,Hong:2023zrf}. 
This indicates that sphalerogenesis does not work within the SM alone, and naturally leads to the question: what kind of new physics can make this mechanism viable?

In Ref.~\cite{Tanaka:2025cpw}, it was shown that sphalerogenesis is feasible in the presence of a dimension-six operator involving the $SU(2)_L$ gauge fields in Eq.~\eqref{eq:EW_Weinberg}, called the EW-Weinberg operator, which acts as a source of CP asymmetry in the sphaleron process. 
In that analysis, the CP asymmetry was estimated by focusing on the difference in sphaleron transition rates along the positive and negative Chern-Simons number directions, instead of relying on sphaleron explosion dynamics. 
While the feasibility of sphalerogenesis was established within an effective field theoretical framework, it remains unclear what ultraviolet-complete theories can realize this scenario.

In this paper, we consider SM extensions with new $SU(2)_L$ multiplet fields~\cite{Banno:2024apv,Banno:2026hsc}, which we refer to as the $SU(2)_L$ multiplet models. 
Such extensions have been widely discussed because the neutral component of a $SU(2)_L$ multiplet can serve as dark matter (DM) candidate~\cite{Cirelli:2005uq,Cirelli:2007xd,Cirelli:2009uv}. 
Furthermore, the sphaleron transition in this type of model has also been extensively analyzed~\cite{Wu:2023mjb, Li:2025kyo}.
In the $SU(2)_L$ multiplet model, Yukawa interactions among the new multiplet fields can contain a CP-violating phase. 
After integrating out the new particles, this CP violation induces the EW-Weinberg operator, which in turn generates CP asymmetry in the EW sphaleron process~\cite{Banno:2024apv}. 
Moreover, the same source of CP violation can be probed by current and future measurements of the electron electric dipole moment (EDM), such as ACME II~\cite{ACME:2018yjb}, JILA~\cite{Roussy:2022cmp}, and ACME III~\cite{Hiramoto:2022fyg}. 
Recently, the electron EDM in the $SU(2)_L$ multiplet models was evaluated at the full three-loop level, and it was shown that these models can be tested in upcoming EDM experiments even when the new particle mass scale is $\mathcal{O}(1)\,\mathrm{TeV}$~\cite{Banno:2026hsc}.

We show that the $SU(2)_L$ multiplet models can provide a viable realization of sphalerogenesis while remaining consistent with the current stringent bound from the electron EDM measurement~\cite{Roussy:2022cmp} as well as LHC search bounds~\cite{Ostdiek:2015aga,Matsumoto:2017vfu,Matsumoto:2018ioi}. 
On the other hand, we find that the conventional freeze-out dark matter scenario is difficult to reconcile with sphalerogenesis, because successful baryogenesis prefers relatively light $SU(2)_L$ multiplet fields. 
This indicates that explaining the baryon asymmetry and dark matter simultaneously in this framework requires a non-freeze-out mechanism for generating the dark matter relic abundance.

The rest of this paper is organized as follows. 
Section~\ref{sec:EWsphaleron_ansatz} introduces the framework for describing the sphaleron process. 
In Section~\ref{sec:CPasymmetry_sph}, we derive the CP asymmetry in sphaleron-like transitions. 
Section~\ref{sec:sphalerogenesis_EWweinberg} presents the sphaleron-like transition rate based on lattice results and formulates the Boltzmann equation for sphalerogenesis, which we numerically solve to estimate the resulting baryon asymmetry. 
In Section~\ref{sec:EWmultiplet}, we introduce the $SU(2)_L$ multiplet model, determine the parameter regions consistent with the observed BAU and current experimental bounds, and discuss the prospects for future EDM measurements and the HL-LHC. 
Section~\ref{sec:conclusion} contains our discussion and conclusions.

\section{Electroweak sphaleron configuration} \label{sec:EWsphaleron_ansatz}

We briefly review the derivation of the sphaleron transition action, which is the basis for estimating the CP asymmetry in the EW sphaleron process. 
We consider the bosonic action
\begin{align}
\label{eq:Stot}
S = S_{\rm SM} + S_{\rm CP} \,. 
\end{align}
The first term is the SM contribution,
\begin{align}
S_{\rm SM} = \int d^4 x \left[ - \frac{1}{2}{\rm tr} \left( W_{\mu \nu} W^{\mu \nu} \right) +|D_{\mu} \Phi|^2 - V(\Phi) \right] \,, 
\end{align}
where $W_{\mu\nu}$ and $\Phi$ denote the $SU(2)_L$ gauge field strength and the SM Higgs doublet, respectively. 
Since the contribution of the $U(1)_Y$ gauge field to the sphaleron process is expected to be below the few-percent level~\cite{Klinkhamer:1984di,Klinkhamer:1990fi}, we neglect it in our analysis. 
The Higgs potential is given by
\begin{align}
V(\Phi) = - \mu^2 |\Phi|^2 + \lambda |\Phi|^4 \,.
\end{align}

The second term in Eq.~\eqref{eq:Stot} encodes the CP-violating effect of new physics. 
Here we work within the Standard Model Effective Field Theory (SMEFT) framework~\cite{Grzadkowski:2010es}. 
As the source of the additional CP violation in Eq.~\eqref{eq:Stot}, we introduce
\begin{align}
\label{eq:EW_Weinberg}
S_{\rm CP} = -  \frac{g}{3 \Lambda^2} \int d^4 x \, f_{ijk} \widetilde{W}^{i}_{\mu \nu} W^{j \nu \rho} W^{k \mu}_{\rho}  \,.
\end{align} 
Here, $g$ and $f_{ijk}$ are the gauge coupling and the structure constants of the $SU(2)_L$ gauge group. 
The dual field strength is defined as $\widetilde{W}^{i}_{\mu \nu}  \equiv \frac{1}{2} \epsilon_{\mu \nu \rho \sigma} W^{i \rho \sigma}$ with $\epsilon_{0123} = 1$. 
This operator, called the EW-Weinberg operator, provides a source of CP violation in the EW sphaleron process~\cite{Tanaka:2025cpw}. 
We note that this operator can naturally arise when $SU(2)_L$ multiplet fields with CP-violating Yukawa interactions are added to the SM~\cite{Banno:2024apv,Banno:2026hsc}. 

We then introduce the ansatz for the EW sphaleron configuration. 
As shown in Ref.~\cite{Manton:1983nd}, the EW sphaleron is described by
\begin{align}
\label{eq:sph_ansatz}
\begin{aligned}
&W_{\mu}(\mu, r, \theta, \phi) dx^{\mu} 
= - \frac{i}{g} f(r) \, d U_{\infty} U^{-1}_{\infty} \,,  \\
&\Phi(\mu , r, \theta, \phi) 
= \frac{v}{\sqrt{2}} \left[ 1- h(r) \right] \left( \begin{array}{c} 0 \\ e^{-i\mu} c_{\mu} \end{array} \right) 
+ \frac{v}{\sqrt{2}} h (r) U_{\infty} \left( \begin{array}{c} 0 \\ 1 \end{array} \right) \,, 
\end{aligned}
\end{align}
with $c_{x} \equiv \cos x$ and $s_{x} \equiv \sin x$. 
The parameters $(r, \theta, \phi)$ and $\mu$ denote the spatial coordinates and the non-contractible loop parameter, respectively. 
The profile functions $f(r)$ and $h(r)$ describe the radial dependence of the $SU(2)_L$ gauge and Higgs fields and satisfy the boundary conditions
\begin{align}
f(0) = h(0) = 0 \,, \quad \lim_{r \to \infty} f(r) =  \lim_{r \to \infty} h(r) = 1 \,.
\end{align}
Their exact forms are obtained by solving the field equations for the $SU(2)_L$ gauge and Higgs fields~(see Refs.~\cite{Klinkhamer:1984di, Gan:2017mcv, Kanemura:2020yyr}). 
Instead of using the numerical solutions directly, we adopt the following approximate profile functions~\cite{Klinkhamer:1984di}:
\begin{align}
\label{eq:fb_hb}
&f(\xi) = \begin{cases}
\frac{\xi^2}{\Xi(\Xi + 4)} \quad & (\xi \leq \Xi) \\ 
1 - \frac{4}{\Xi + 4} \exp \left[ \frac{\Xi - \xi}{2} \right] & (\xi > \Xi) \,, 
\end{cases}  \\ 
&h(\xi) = \begin{cases}
\frac{\sigma \Omega + 1}{\sigma \Omega +2} \frac{\xi}{\Omega} \quad & (\xi \leq \Omega) \\ 
1 - \frac{\Omega}{\sigma \Omega + 2} \frac{1}{\xi} \exp \left[ \sigma (\Omega - \xi) \right] & (\xi > \Omega) \,, 
\end{cases} 
\end{align}
with $\sigma = \sqrt{2\lambda/g^2}$. 
In our analysis, we use $m_{h} = 125.1\,{\rm GeV}$ and $m_{W} = 80.329\,{\rm GeV}$ for the Higgs and $SU(2)_L$ gauge boson masses, respectively~\cite{ParticleDataGroup:2024cfk}. 
The parameters $\Xi$ and $\Omega$ are determined by requiring that the bosonic energy functional be minimized at $\mu = \pi/2$, which corresponds to the saddle point between adjacent topologically distinct vacua~\cite{Manton:1983nd,Klinkhamer:1984di}. 
We numerically confirm that the energy functional is minimized for
\begin{align}
\label{eq:Xi0_Omega0}
\Xi = 1.467 \equiv \Xi_{0} \,, \quad \Omega = 1.701 \equiv \Omega_{0} \,, 
\end{align}
in agreement with Ref.~\cite{Hong:2023zrf}. 
In the following, we refer to the configuration with Eq.~\eqref{eq:Xi0_Omega0} as the true sphaleron. 
The $2 \times 2$ matrix $U_{\infty}$ is given by~\cite{Manton:1983nd}
\begin{align}
  U_{\infty} =
  \left( \begin{array}{cc} 
  e^{i \mu} (c_{\mu} - i s_{\mu} c_{\theta}) & e^{i \phi} s_{\mu} s_{\theta} \\
  - e^{-i \phi} s_{\mu} s_{\theta} & e^{-i\mu} (c_{\mu} + i s_{\mu} c_{\theta})
  \end{array} \right) \,.
\end{align}
For the vacuum expectation value of the SM Higgs field $v$ during the smooth EW symmetry breaking, we use the fitting formula obtained from lattice simulations~\cite{DOnofrio:2014rug,Kharzeev:2019rsy}
\begin{align}
v(T) \simeq 3 T \sqrt{1 - \frac{T}{T_{\rm EW}}} \,,
\end{align}
where $T_{\rm EW}$ is the crossover temperature, $T_{\rm EW} \simeq 159.5\,{\rm GeV}$, in the SM~\cite{DOnofrio:2014rug, DOnofrio:2015gop, Annala:2023jvr}. 

Substituting Eq.~\eqref{eq:sph_ansatz} into Eq.~\eqref{eq:Stot}, we obtain the sphaleron transition action
\begin{align}
\label{eq:Ssph}
S_{\rm sph} = \int d t \left[ \frac{M(\mu)}{2} \left( \frac{d \mu}{d \eta} \right)^2 + G(\mu) \left( \frac{d\mu}{d \eta}\right)^3 - V(\mu) \right] \,, 
\end{align}
with $\eta = g v t$. 
The explicit expressions for $M(\mu)$, $G(\mu)$, and $V(\mu)$ are given in Appendix~\ref{sec:sphaleron_action}. 
In deriving Eq.~\eqref{eq:Ssph}, we used the scheme suggested in Ref.~\cite{Funakubo:2016xgd}. 
Using the profile functions in Eq.~\eqref{eq:fb_hb} with the parameters in Eq.~\eqref{eq:Xi0_Omega0}, the action in Eq.~\eqref{eq:Ssph} can be written as~\cite{Nauta:2002ru,Funakubo:2016xgd,Tanaka:2025cpw}
\begin{align}
&M(\mu) = \frac{4\pi v(T)}{g}  \left( \alpha_{0} + \alpha_{1} c_{\mu}^2 + \alpha_{2} c_{\mu}^4 \right) \,, \\
&G(\mu) = \frac{256 \pi}{45} g v(T) s_{\mu}^2 (4 - s_{\mu}^2) \left( \frac{v(T)}{\Lambda} \right)^2 \,, \\
&V(\mu) = \frac{4 \pi v(T)}{g} s_{\mu}^2 \left( \beta_{1} + \beta_{2} s_{\mu}^2 \right) \,,
\end{align}
where $\alpha_{i}\,(i=0,1,2)$ and $\beta_{j}\,(j=1,2)$ are determined by $f(\xi)$ and $h(\xi)$
\footnote{
Other CP-violating dimension-six operators, except for the EW-Weinberg operator, do not generate the cubic term in $S_{\rm sph}$. For example, for the operator $|\Phi|^2 {\rm tr}[W_{\mu \nu} \widetilde{W}^{\mu \nu}]$, one obtains only a total time derivative term in the sphaleron action $S_{\rm sph}$~\cite{Tanaka:2025cpw}. 
Therefore, it does not affect the sphaleron dynamics.}. 
Using Eqs.~\eqref{eq:fb_hb} and \eqref{eq:Xi0_Omega0}, we obtain
\begin{align}
\label{eq:sph_parameters}
\alpha_{0} = 19.6 \,, \quad \alpha_{1} = -1.84 \,, \quad \alpha_{2} = -2.64 \,, \quad  \beta_{1} = 1.37 \,, \quad \beta_{2} = 0.57 \,. 
\end{align}

Eq.~\eqref{eq:Ssph} shows that the sphaleron action can be interpreted as the effective action for a $1+1$-dimensional dynamical system, where the loop parameter $\mu$ is promoted to a dynamical variable through $Q =\mu/(g v)$. 
This effective description of the sphaleron transition is known as the reduced model~\cite{Aoyama:1987nd,Funakubo:1991hm,Funakubo:1992nq,Nauta:2000xi,Nauta:2002ru,Tye:2015tva,Tye:2016pxi,Tye:2017hfv,Funakubo:2016xgd,Qiu:2018wfb}. 
We use this reduced model to estimate the CP asymmetry in the EW sphaleron process. 

In the early universe, thermal fluctuations may cause vacuum transitions to occur along paths that deviate from the least energy path.
Such sphaleron-like transitions proceed through sphaleron-like field configurations instead of the true sphaleron.
To describe sphaleron-like configurations, we parametrize deviations of the profile functions as
\begin{align}
\label{eq:fh_alpha_beta}
\left. f(\xi) \right|_{\Xi_{0} \to \alpha \Xi_{0}} \,, \quad  \left. h(\xi) \right|_{\Omega_{0} \to \beta \Omega_{0}} \,. 
\end{align}
In previous works~\cite{Hong:2023zrf,Tanaka:2025cpw}, these deformations were parameterized by a single parameter, $\alpha = \beta = a$, for simplicity. 
Here, in contrast, we introduce two independent parameters, $\alpha$ and $\beta$. 
As we show later, this generalization leads to a slight enhancement of the produced baryon asymmetry. 
With the replacement in Eq.~\eqref{eq:fh_alpha_beta}, the action for sphaleron-like processes is obtained by
\begin{align}
\label{eq:replace_MGV}
M(Q) \to M(\alpha, \beta, Q) \,, \quad 
G(Q) \to G(\alpha, \beta, Q) \,, \quad 
V(Q) \to V(\alpha, \beta, Q) \,. 
\end{align}

\section{Estimation of CP asymmetry in EW sphaleron transition} \label{sec:CPasymmetry_sph}

We then estimate the CP asymmetry in the EW sphaleron process using the effective action introduced in the previous section. 
According to Eq.~\eqref{eq:Ssph}, the corresponding effective Lagrangian in terms of $Q$ is
\begin{align}
L = \frac{M(Q)}{2} \left( \frac{dQ}{dt} \right)^2 + G(Q) \left( \frac{dQ}{dt} \right)^3 - V(Q) \,. 
\end{align}
In order to move on to the Hamilton formalism, we define the canonical conjugate momentum of $Q$ as
\begin{align}
\pi_{Q} = \frac{\partial L}{\partial \dot{Q}} = M(Q) \dot{Q} + 3 G(Q) \dot{Q}^2 \,. 
\end{align}
The corresponding Hamiltonian is then given by
\begin{align}
\label{eq:Hamiltonian_sph}
\mathcal{H} = \frac{1}{2M(Q)} \pi_{Q}^2 - \frac{G(Q)}{M^3(Q)} \pi_{Q}^3 + V(Q) \,. 
\end{align}
Because the effective Hamiltonian contains the cubic term $\pi_{Q}^3$, the kinematics for positive and negative $\pi_{Q}$ are no longer identical. 
Equivalently, the cubic term induces a difference between the sphaleron transition rates toward positive and negative Chern-Simons numbers. 
This asymmetry is the origin of CP violation in the EW sphaleron process. 

To quantify the CP asymmetry relevant for baryogenesis, we focus on the probability current associated with the sphaleron-like transition. 
This is different from the equilibrium diffusion discussed in Refs.~\cite{Nauta:2000xi,Nauta:2002ru}, where the net probability current vanishes. 
When sphaleron-like configurations are decoupled from the thermal bath, their subsequent evolution is no longer balanced by the inverse process. 
Thus, it is natural to characterize the CP asymmetry by the difference between the probability currents along the positive and negative $Q$ directions.
We define the corresponding probability current at the saddle-point $\pm Q_{\rm sph} = \pm \pi/(2g v)$ along the positive and negative $Q$ directions by~\cite{Affleck:1980ac}
\begin{align}
\label{eq:Gpm_flux}
\Gamma_{\pm}(Q_{\rm sph})
=
\frac{1}{Z_{0}}
\iint d\pi_Q\, dQ\;
\pi_Q\, e^{-\mathcal H(Q,\pi_Q)/T}\,
\Theta(\pm \pi_Q)\,
\delta(Q \mp Q_{\rm sph}) \,,
\end{align}
where $Z_{0}$ is a normalization factor. 
$\Theta(x)$ and $\delta(x)$ denote the step and Dirac delta functions, respectively. 
The step function is inserted to remove the contribution of inverse sphaleron-like transitions.
Evaluating the integral for $Q$ and expanding to the linear order in the cubic term because of $G(Q_{\rm sph}) \ll T, \,M(Q_{\rm sph})$, one finds
\begin{align}
\Gamma_{\pm}(Q_{\rm sph})
\simeq
\frac{1}{Z_{0}}
e^{-V(Q_{\rm sph})/T}
\int_{0}^{\infty} d\pi_Q\,
\pi_Q\,
\left(
1 \pm \frac{G(Q_{\rm sph})}{M(Q_{\rm sph})^3} \frac{\pi_Q^3}{T}
\right)
\exp \left[-\frac{\pi_Q^2}{2M(Q_{\rm sph})T}\right] \,.
\end{align}
One may be concerned with the dynamical instability of the system due to the cubic term in Eq.~\eqref{eq:Hamiltonian_sph} at the large $\pi_{Q}$ region. 
However, since the thermal averaged $\pi_{Q}$ is roughly given by $\braket{\pi_{Q}} \sim \sqrt{TM(Q_{\rm sph})}$, the second term in Eq.~\eqref{eq:Hamiltonian_sph} is significantly smaller than the first term. 
Therefore, we expect that such a dynamical instability does not appear, and the above approximation is plausible. 

We here assume that the CP asymmetry in EW sphaleron processes can be quantified by
\begin{align}
A_{\rm CP}
\equiv
\frac{\Gamma_+(Q_{\rm sph})-\Gamma_-(Q_{\rm sph})}
{\Gamma_+(Q_{\rm sph})+\Gamma_-(Q_{\rm sph})} \,. 
\end{align}
Using Eq.~\eqref{eq:Gpm_flux} and performing Gauss integrals, we obtain
\begin{align}
A_{\rm CP}
=
\frac{3 \pi}{4} 
\sqrt{\frac{8T}{\pi}}
\frac{G(Q_{\rm sph})}{M^{3/2}(Q_{\rm sph})} \,.
\end{align}
Compared with the result in Ref.~\cite{Tanaka:2025cpw}, the CP asymmetry in the EW sphaleron processes is enhanced by the additional factor $3\pi/4$.\footnote{In the previous work~\cite{Tanaka:2025cpw}, $M(Q)$ and $G(Q)$ are normalized by $gv(T)$.}

Our derivation of the effective CP asymmetry relevant for baryogenesis is subject to several sources of theoretical uncertainty. 
In particular, our results include uncertainties related to the reduced model, such as the treatment of the effective kinetic term $M(\mu)$~\cite{Funakubo:2016xgd, Tye:2015tva}, the use of time-independent profile functions $f(\xi)$ and $h(\xi)$ (see Ref.~\cite{Matchev:2025irm}), and the limitation of the reduced model description itself. 
Some of these uncertainties can already modify the predicted asymmetry at the level of a factor of order unity. 
For example, different reduced constructions of the sphaleron action lead to different effective masses $M(Q)$, which in turn induce an $O(1)$ variation in the CP asymmetry because of $A_{\rm CP} \propto M^{-3/2}(Q)$~\cite{Funakubo:2016xgd, Tye:2015tva}. 
A fully systematic treatment of all these effects requires a time-dependent and multidimensional analysis of the sphaleron decoupling dynamics, which is beyond the scope of the present work.
Instead, we absorb their effects into a single phenomenological factor $\kappa_{\rm CP}$, and write the effective CP asymmetry in sphaleron processes as
\begin{align}
\label{eq:Acp_sph}
A_{\rm CP}^{\rm eff} = \kappa_{\rm CP} A_{\rm CP} \,.
\end{align}
In the following, we assume that $A_{\rm CP}^{\rm eff}$ quantifies the CP asymmetry in the sphaleron-like transition entering the source term of the Boltzmann equation. 
For the value of $\kappa_{\rm CP}$, we mainly focus on the following region in this paper
\begin{align}
\label{eq:kappa_range}
0.5 \leq \kappa_{\rm CP} \leq 3 \,.
\end{align}

\section{Sphalerogenesis with the EW-Weinberg operator} \label{sec:sphalerogenesis_EWweinberg}

\subsection{Sphaleron-like transition rate}

In this section, we briefly review the mechanism of sphalerogenesis discussed in Ref.~\cite{Tanaka:2025cpw}. 
In this scenario, the sphaleron-like transition rate plays a central role. 
Following Ref.~\cite{Hong:2023zrf}, we define the sphaleron-like transition rate by
\begin{align}
\Gamma_{\rm sph}^{\rm lattice}(T) = \int d \alpha \, d \beta \, \Gamma_{\rm sph}(\alpha, \beta, T) \,,
\end{align}
with
\begin{align}
\label{eq:Gamma_sph_a_b}
\Gamma_{\rm sph}(\alpha, \beta, T) = 
\frac{e^{-E_{\rm sph}(\alpha, \beta, T)/T}}
{\int d \alpha' \, d\beta' \, e^{-E_{\rm sph}(\alpha', \beta', T)/T}}
\, \Gamma_{\rm sph}^{\rm lattice}(T) \,.
\end{align}
Here $E_{\rm sph}(\alpha, \beta, T)$ denotes the energy of the sphaleron-like configuration: $E_{\rm sph}(\alpha, \beta, T) =  V(\alpha, \beta, \mu=\pi/2)$. 
The quantity $\Gamma_{\rm sph}^{\rm lattice}(T)$ is the total sphaleron transition rate obtained from lattice simulations~\cite{DOnofrio:2014rug,Annala:2023jvr}. 
In the broken phase ($T < T_{\rm EW}$), the fitting function for the sphaleron transition rate is given by~\cite{Annala:2023jvr}
\begin{align}
\label{eq:Gamma_lattice}
\log \left[ \frac{\Gamma_{\rm sph}^{\rm lattice}(T)}{T^4} \right] = \left( 0.86 \pm 0.01 \right) \frac{T}{{\rm GeV}} - \left( 153.1 \pm 0.9 \right)  \,. 
% \log \left[ \frac{\Gamma_{\rm sph}^{\rm lattice}(T)}{T^4} \right] =  X \left( \frac{T}{{\rm GeV}} \right) + Y  \,. 
\end{align}
In the following analysis, we estimate the resulting baryon asymmetry by using $\Gamma_{\rm sph}^{\rm lattice}(T)$ with the central value in Eq.~\eqref{eq:Gamma_lattice}.

The sphaleron decoupling temperature $T_{\rm sph}$ is commonly defined by
\begin{align}
\label{eq:sph_decoupling_lattice}
\frac{\Gamma_{\rm sph}^{\rm lattice}(T_{\rm sph})}{T_{\rm sph}^3}  = k H(T_{\rm sph}) \,,
\end{align}
where $k = 4/39 \simeq 0.1$ is conventionally adopted as a representative value~\cite{Bochkarev:1987wf,DOnofrio:2014rug,Annala:2023jvr}. 
Using Eq.~\eqref{eq:Gamma_lattice}, one obtains $T_{\rm sph} = 133.5 \pm 0.97\,{\rm GeV}$~\cite{Annala:2023jvr}, which is slightly higher than the previous result $T_{\rm sph} = 131.7 \pm 2.3\,{\rm GeV}$ reported in Ref.~\cite{DOnofrio:2014rug}. 

As the temperature decreases, sphaleron-like processes with $(\alpha,\beta)\neq(1,1)$ decouple earlier than the true sphaleron, because their rates are more strongly Boltzmann suppressed: $E_{\rm sph}(\alpha,\beta,T)\ge E_{\rm sph}(1,1,T)$. 
By analogy with Eq.~\eqref{eq:sph_decoupling_lattice}, we determine the decoupling temperature of each sphaleron-like process through~\cite{Hong:2023zrf,Tanaka:2025cpw}
\begin{align}
\label{eq:sph_decoupling}
\frac{\Gamma_{\rm sph}(\alpha , \beta, T)}{T^3} = c H(T) \,,
\end{align}
where $H(T)$ is the Hubble parameter. 
In this framework, we can determine the temperature $\widetilde{T}_{\rm sph}$ at which all sphaleron-like transitions decouple as follows
\begin{align}
\frac{\Gamma_{\rm sph}(\alpha = 1 , \beta = 1, \widetilde{T}_{\rm sph})}{\widetilde{T}_{\rm sph}^3} = c H(\widetilde{T}_{\rm sph}) \,,
\end{align}
where the factor $c$ is determined by requiring $T_{\rm sph} = \widetilde{T}_{\rm sph}$. 
We have numerically obtained $c \simeq 0.28$. 
The discrepancy between $k$ and $c$ is due to the prefactor in Eq.~\eqref{eq:Gamma_sph_a_b}, which makes $\Gamma_{\rm sph}(\alpha, \beta, T)$ smaller than $\Gamma_{\rm sph}^{\rm lattice}(T)$. 
As a result, we obtain $\widetilde{T}_{\rm sph} \geq T_{\rm sph}$ if $k = c$. 

\begin{figure*}[t]
    \centering
    \includegraphics[width=0.98\linewidth]{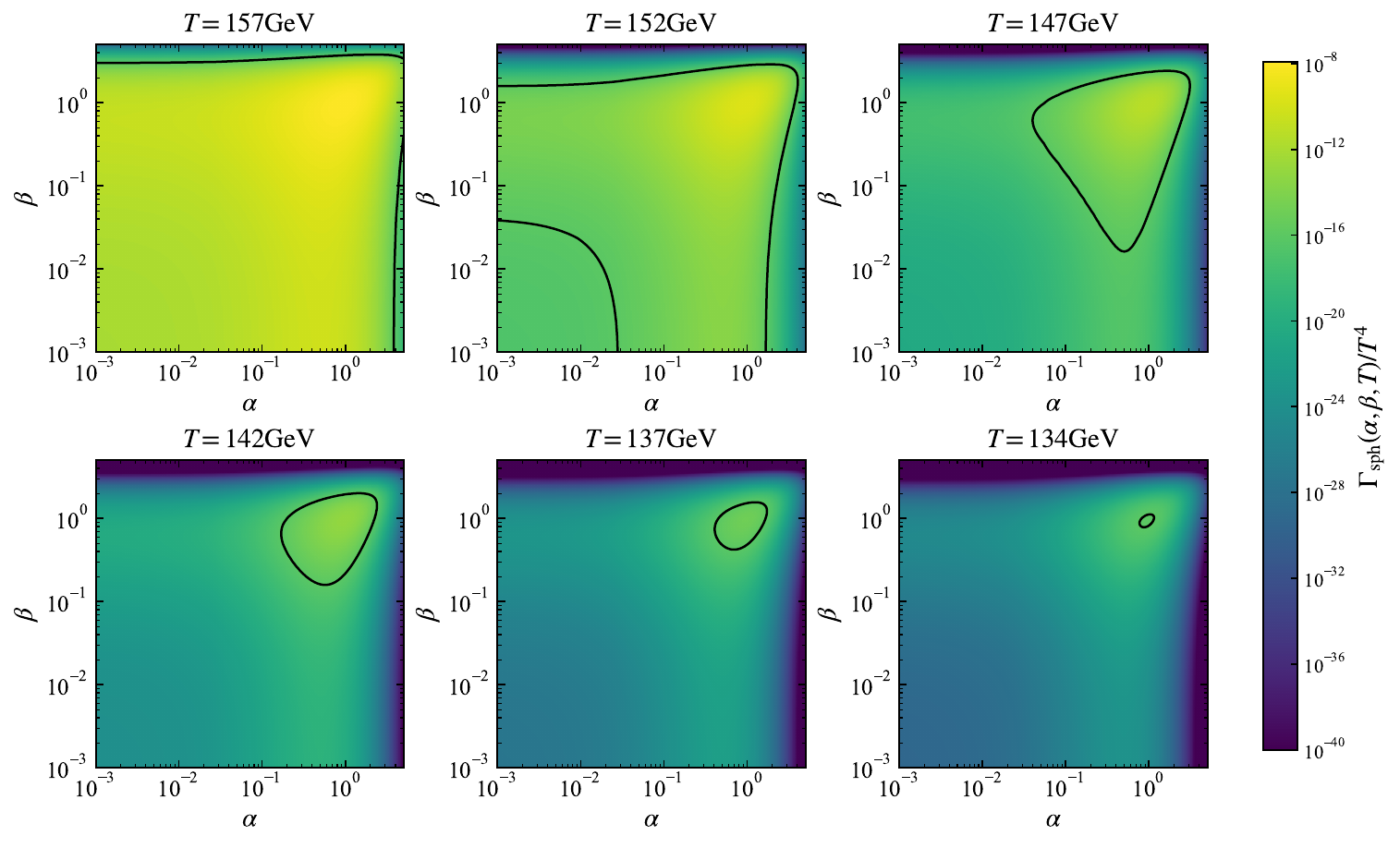}
    \caption{
    Temperature dependence of the sphaleron-like transition rate in the $(\alpha,\beta)$ plane. 
    The black solid contours show the boundary defined by Eq.~\eqref{eq:sph_decoupling}. 
    Sphaleron-like transitions inside (outside) the contours are active (inactive). 
    }
    \label{fig:Gamma_sph_ab}
\end{figure*}

Figure~\ref{fig:Gamma_sph_ab} shows the temperature dependence of the sphaleron-like transition rate in the $(\alpha,\beta)$ plane. 
As the temperature decreases, the active region gradually shrinks. 
During this decoupling process, sphaleron-like configurations cease to be produced thermally and eventually disappear. 
This gradual decoupling provides the departure from thermal equilibrium required for baryogenesis~\cite{Sakharov:1967dj}. 
Below $T \simeq 133\,{\rm GeV}$, all sphaleron-like processes are decoupled. 
In the following, the region inside (outside) the contour in Fig.~\ref{fig:Gamma_sph_ab} is denoted by $S(T)$ ($\overline{S}(T)$).

\subsection{Boltzmann equation for sphalerogenesis}

As we have seen in Fig.~\ref{fig:Gamma_sph_ab}, the decay and formation of sphaleron-like configurations in the region $S(T)$ occur frequently. 
Therefore, they mainly contribute to the washout of the produced baryon asymmetry. 
On the other hand, configurations in the region $\overline{S}(T)$ can still decay efficiently, but they are no longer reproduced from the thermal bath because their production processes have already decoupled. 
If these decays are CP asymmetric, a net baryon asymmetry is generated during the decoupling of the sphaleron-like transitions. 
Using $A_{\rm CP}^{\rm eff}$ in Eq.~\eqref{eq:Acp_sph}, the Boltzmann equation for the baryon asymmetry $n_{B}$ is given by~\cite{Tanaka:2025cpw}
\begin{align}
\label{eq:BoltzmannEq}
-HT \frac{d n_{B}}{dT} + 3 H n_{B} = - \Gamma_{B}(T) n_{B} + P(T) \,,
\end{align}
where $\Gamma_{B}(T)$ and $P(T)$ denote the washout and source terms, respectively. 
Their explicit forms are
\begin{align}
\Gamma_{B}(T) = 
\begin{cases}
\displaystyle \frac{39}{4T^3} \int_{S(T)} d \alpha \, d \beta \, \Gamma_{\rm sph} (\alpha, \beta, T)
& (T_{\rm sph} < T < T_{\rm EW}) \,, \\[2mm]
0 & (T < T_{\rm sph}) \,,
\end{cases}
\end{align}
and
\begin{align}
P(T) = 
\begin{cases}
\displaystyle \int_{\overline{S}(T)} d \alpha \, d \beta \,  \Gamma_{\rm sph}(\alpha, \beta, T) \cdot 3 A_{\rm CP}^{\rm eff}(\alpha, \beta, T)
& (T_{\rm sph} < T < T_{\rm EW}) \,, \\ 
\Gamma_{\rm sph}^{\rm lattice}(T) \cdot 3 A_{\rm CP}^{\rm eff} (\alpha = 1, \beta = 1, T)
& (T < T_{\rm sph}) \,.
\end{cases}
\end{align}
For $T<T_{\rm sph}$, we approximate the source term by the contribution from the final decay of the true sphaleron configuration. 
We note that the washout term integrates over active configurations in $S(T)$, whereas the source term integrates over decoupled configurations in $\overline{S}(T)$. 
The factor 3 in $P(T)$ reflects the net baryon number change $\Delta B = 3$ induced by a single electroweak sphaleron transition~\cite{tHooft:1976snw}. 

\begin{table}[t]
 \begin{center}
   \begin{tabular}{|c|c|} 
    \hline
    ACME~II (2018) \cite{ACME:2018yjb} & $|d_e| < 1.1 \times 10^{-29}\, e\,{\rm cm}$ \\ 
    \hline
    JILA (2023) \cite{Roussy:2022cmp} & $|d_e| < 4.1 \times 10^{-30}\, e \, {\rm cm}$ \\ 
    \hline
    ACME~III (future) \cite{Hiramoto:2022fyg} & $|d_e| < \mathcal{O}(10^{-31})\, e \, {\rm cm}$ \\ 
    \hline
  \end{tabular}
   \caption{Current and projected upper bounds from electron EDM experiments. \label{tab:EDMconstraint}}
 \end{center}
\end{table}

Sphalerogenesis is strongly constrained by electron EDM experiments. 
The sensitivity of these experiments has improved dramatically through the use of polar molecules. 
Table~\ref{tab:EDMconstraint} summarizes the current and projected bounds. 
When the operator in Eq.~\eqref{eq:EW_Weinberg} is added to the SM, the induced electron EDM is given by~\cite{Boudjema:1990dv,Banno:2024apv}
\begin{align}
\label{eq:EDM_EFT}
\frac{d_{e}^{\rm EFT}}{e} = \frac{g^2 m_{e}}{96 \pi^2 \Lambda^2} \simeq 4.1 \times 10^{-30}\,{\rm cm} \left( \frac{33\,{\rm TeV}}{\Lambda}\right)^2\,,
\end{align}
where $m_{e}$ is the electron mass. 
In this section, we regard Eq.~\eqref{eq:EDM_EFT} as the prediction for the electron EDM within the SMEFT framework. 

\begin{figure*}[t]
\centering
\subfigure[Solution of the Boltzmann equation in Eq.~\eqref{eq:BoltzmannEq}.\label{fig:nbs_T}]{\includegraphics[width=0.487\linewidth]{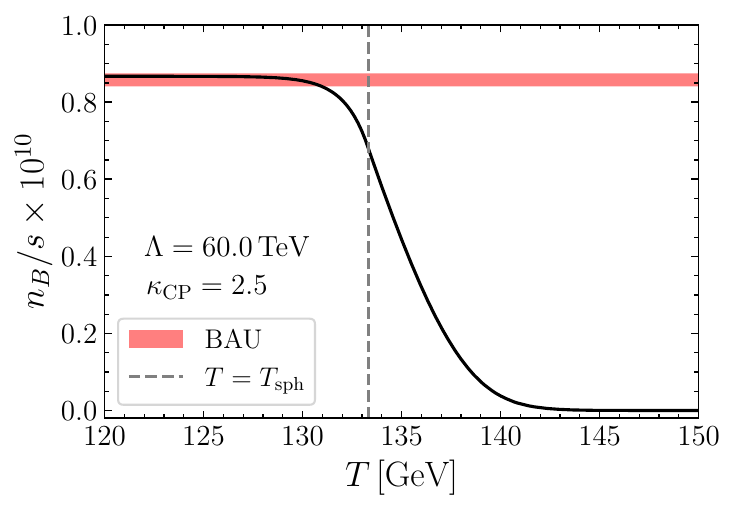}}
\subfigure[Cutoff dependence of $n_{B}/s$ for each $\kappa_{\rm CP}$.\label{fig:bau_Lambda}]{\includegraphics[width=0.48\linewidth]{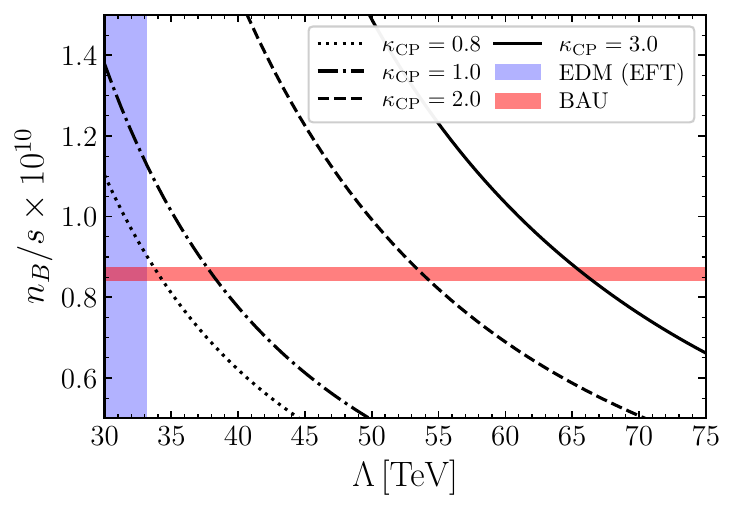}}
\caption{
\textbf{[Left]}
Temperature evolution of $n_{B}/s$ for $\Lambda = 60\,{\rm TeV}$ and $\kappa_{\rm CP} = 2.5$.
The red band indicates the observed value of $n_{B}/s$ in Eq.~\eqref{eq:BAU}. 
The gray dashed line marks the sphaleron decoupling temperature $T_{\rm sph} = 133.3\,{\rm GeV}$. 
\textbf{[Right]}
Cutoff-scale dependence of the baryon asymmetry for each value of $\kappa_{\rm CP}$.
The blue region is excluded by the current electron EDM bound from JILA~\cite{Roussy:2022cmp}, using Eq.~\eqref{eq:EDM_EFT}. 
For the right side of the black solid line, Eq.~\eqref{eq:kappa_range} is not satisfied. 
}
\label{fig:bau_EFT}
\end{figure*}

Figure~\ref{fig:nbs_T} shows a numerical solution of the Boltzmann equation for $\Lambda = 60\,{\rm TeV}$ and $\kappa_{\rm CP} = 2.5$. 
The red band indicates the observed baryon asymmetry in Eq.~\eqref{eq:BAU}, and the gray dashed line marks the sphaleron decoupling temperature $T_{\rm sph}$. 
As seen in the figure, the baryon asymmetry is generated predominantly around $T \sim T_{\rm sph}$ because of $A_{\rm CP}^{\rm eff} \propto v(T)^{3/2}$. 
Figure~\ref{fig:bau_Lambda} shows the cutoff dependence of $n_{B}/s$ for several values of $\kappa_{\rm CP}$. 
The blue region is excluded by the current strongest electron EDM bound~\cite{Roussy:2022cmp} using Eq.~\eqref{eq:EDM_EFT}. 
When we scan $\kappa_{\rm CP}$ within the range in Eq.~\eqref{eq:kappa_range}, the observed BAU can be reproduced without violating the current EDM bound if the cutoff scale lies in the range
\begin{align}
\label{eq:BAULambda_EFT}
33.1\,{\rm TeV} < \Lambda < 66.6\,{\rm TeV} \,.
\end{align}
For the value of $\kappa_{\rm CP}$, it should be larger than 0.8 to explain the BAU. 
In the next section, we use Eq.~\eqref{eq:BAULambda_EFT} to identify parameter regions of the $SU(2)_L$ multiplet models in which the BAU can be explained. 

Furthermore, we have confirmed that the required value of $\Lambda$ to explain the BAU is hardly changed from the result in Ref.~\cite{Tanaka:2025cpw} even if we introduce two independent parameters $\alpha$ and $\beta$ to characterize the sphaleron-like field configuration. 
This is because the baryon asymmetry is generated mainly around $T_{\rm sph}$, where the active region $S(T)$ is already narrow, as shown in Fig.~\ref{fig:Gamma_sph_ab}. 
Therefore, the approximation $\alpha=\beta$ used in Refs.~\cite{Hong:2023zrf, Tanaka:2025cpw} already captures the essential features of sphalerogenesis.

We emphasize that the electron EDM bound in Fig.~\ref{fig:bau_Lambda} depends sensitively on the ultraviolet completion. 
In Fig.~\ref{fig:bau_Lambda}, we used the relation between the EW-Weinberg operator in Eq.~\eqref{eq:EW_Weinberg} and the electron EDM prediction in Eq.~\eqref{eq:EDM_EFT}. 
However, Ref.~\cite{Banno:2026hsc} showed that the electron EDM in the $SU(2)_L$ multiplet models receives an additional threshold contribution at the three-loop level. 
As a result, in the heavy new particle limit, one finds $d_{e}^{\rm Full} \simeq 3\,d_{e}^{\rm EFT}$. 
Therefore, the electron EDM constraint in $SU(2)_L$ multiplet models is substantially stronger than the one shown in Fig.~\ref{fig:bau_Lambda}. 
We discuss this point in detail in Section~\ref{sec:EWmultiplet}.

\section{Application to the $SU(2)_{L}$ multiplet models} \label{sec:EWmultiplet}

\subsection{EW-Weinberg operator in the $SU(2)_{L}$ multiplet models}

We now discuss the $SU(2)_{L}$ multiplet model as a representaive ultraviolet-complete model that generates the EW-Weinberg operator in Eq.~\eqref{eq:EW_Weinberg} at low energies. 
The following discussion is based on Refs.~\cite{Banno:2024apv,Banno:2026hsc}. 

We consider SM extensions with new $SU(2)_L$ multiplet fields, $\psi_A$, $\psi_B$, and $S$, together with CP-violating Yukawa interactions. 
The relevant interaction terms are
\begin{align}
   \mathcal{L} \ni
   - \bar{\psi}_B g_{\bar{B} A S} \psi_A S - \bar{\psi}_A g_{\bar{A} B \bar{S}} \psi_B S^*\,,
    \label{eq:Yukawa int} 
\end{align}
where $\psi_A$ and $\psi_B$ are fermions, and $S$ is a complex scalar, each transforming in each representation of the $SU(2)_L$ gauge group. 
The Yukawa couplings are written as
\begin{align}
    g_{\bar{B}AS} &= X_{\bar{B}AS}(s+\gamma_5a)\,, \\
    g_{\bar{A}B\bar{S}} &= X_{\bar{A}B\bar{S}}(s^*-\gamma_5a^*)\,,
\end{align}
where $s$ and $a$ are in general complex parameters. 
The factors $X_{\bar{B}AS}$ and $X_{\bar{A}B\bar{S}}$ are $SU(2)_L$-invariant tensors determined by the representations of the new fields, and their explicit forms are given in Ref.~\cite{Banno:2024apv}. 
This setup generates the EW-Weinberg operator at the two-loop level~\cite{Banno:2024apv,Abe:2017sam}. 

For simplicity, we focus on the case in which the new fields transform as $(A,B,S)=(r,r,1)$ under the $SU(2)_L$ gauge group, where $r$ denotes an arbitrary representation. 
In this case, $X_{\bar{B}AS}$ and $X_{\bar{A}B\bar{S}}$ reduce to Kronecker delta $\delta^{a_r b_r}$, where its indices are related to the representation of the fermions $A$ and $B$. 
Then the cutoff scale $\Lambda$ in Eq.~\eqref{eq:EW_Weinberg} is given by~\cite{Abe:2017sam,Banno:2024apv}
\begin{align}
    \frac{1}{\Lambda^2} = \frac{\alpha_2\, r(r^2 - 1)}{128 \pi^3}
    \Im(sa^*) \, m_A m_B \, F(m_A^2,m_B^2,m_S^2)\,,    
    \label{eq:Wilson coeffiecient}
\end{align}
with
\begin{align}
F(m_A^2,m_B^2,m_S^2) \equiv g_1(m_A^2,m_B^2,m_S^2) + g_2(m_A^2,m_B^2,m_S^2) + (m_A \leftrightarrow m_B) \,,
\end{align}
where $\alpha_2 = g^2/(4\pi)$ and $(m_A \leftrightarrow m_B)$ denotes the exchange of the two fermion masses in $g_1$ and $g_2$. 
The functions $g_1$ and $g_2$ are combinations of two-loop functions:
\begin{align}
    g_1(x_1,x_2,x_3) &= \left( 2\bar{I}_{(4;1)} + 4x_1\bar{I}_{(5;1)} \right)(x_1;x_2;x_3)\,, \\
    g_2(x_1,x_2,x_3) &= \left( \bar{I}_{(3;2)} + x_1\bar{I}_{(4;2)} \right)(x_1;x_2;x_3)\,,
\end{align}
where $\bar{I}_{(n;m)}$ is defined in terms of the finite master integral $\bar{I}(x_1;x_2;x_3)$ as
\begin{align}
    \bar{I}_{(n;m)}(x_1;x_2;x_3)
    = \frac{1}{(n-1)!(m-1)!} \frac{d^{n-1}}{dx_1^{n-1}} \frac{d^{m-1}}{dx_2^{m-1}} \bar{I}(x_1;x_2;x_3)\,.
\end{align}
Analytic expressions for $\bar{I}(x_1;x_2;x_3)$ can be found in Refs.~\cite{Ford:1992pn,Espinosa:2000df,Martin:2001vx}. 
The factor $r(r^2-1)/12$ in Eq.~\eqref{eq:Wilson coeffiecient} arises from the $SU(2)_L$ group structure,
\begin{align}
 \Tr \left[T^3 [T^+ , T^-]\right]=\frac{r (r^2 - 1)}{12}\,,
 \label{eq:factor_r}
\end{align}
where $T^a~(a=1,2,3)$ are the generators of $SU(2)_L$ and $T^\pm = T^1 \pm i T^2$. 
In general, CP-violating observables must be invariant under field redefinitions. 
Eq.~\eqref{eq:Wilson coeffiecient} respects this property because the combination $\Im(sa^*) m_A m_B$, arising from the $\gamma_5$ structure in the fermion trace together with two chirality-flipping mass insertions, is rephasing invariant~\cite{Banno:2026hsc,Banno:2023yrd}. 

Combining Eq.~\eqref{eq:Wilson coeffiecient} with Eq.~\eqref{eq:BAULambda_EFT}, we can identify the parameter region capable of explaining the observed BAU. 
Before doing so, however, we first discuss the electron EDM constraint in this model. 

\subsection{Electron electric dipole moment with $SU(2)_{L}$ multiplets} 

We briefly review the full-loop result for the electron EDM in the $SU(2)_L$ multiplet models, following Ref.~\cite{Banno:2026hsc}. 

The electron EDM induced through the EW-Weinberg operator can be estimated by combining Eqs.~\eqref{eq:EDM_EFT} and \eqref{eq:Wilson coeffiecient}. 
An important point is that the matching between the EW-Weinberg operator and the electron EDM does not receive logarithmic enhancement from renormalization-group running, because the corresponding anomalous dimension vanishes~\cite{Jenkins:2017dyc}. 
Therefore, this contribution should be regarded as a threshold correction in the low-energy effective theory below the EW scale, and it enters at the same order as the lepton dipole operator treated within the SMEFT framework.

In Ref.~\cite{Banno:2026hsc}, the electron EDM induced by the CP-violating Yukawa interactions in Eq.~\eqref{eq:Yukawa int} was evaluated at the full three-loop level, including the contribution of the lepton dipole operator. 
The resulting EDM is
\begin{align}
    \frac{d^{\rm Full}_e}{e}
    =
    \frac{\alpha_2^2}{(16 \pi^2)^2}
    \frac{m_e}{2}
    \frac{r (r^2-1)}{12}
    \Im (s a^*)
    m_A m_B  
    B(m_A,m_B,m_S,m_W)\,,
    \label{eq:eEDM_Full}
\end{align}
where $B(m_A,m_B,m_S,m_W)$ denotes a combination of three-loop integrals. 
Eq.~\eqref{eq:eEDM_Full} is invariant under field redefinitions for the same reason as Eq.~\eqref{eq:Wilson coeffiecient}. 
When the $SU(2)_L$ multiplet fields are heavier than the $W$ boson, $B(m_A,m_B,m_S,m_W)$ can be expanded as
\begin{align}
    B(m_A,m_B,m_S,m_W)
    \simeq
    B_0(m_A,m_B,m_S)
    +
    m^2_W B_1(m_A,m_B,m_S)
    +
    \mathcal{O} \left( \frac{m^4_W}{M^4} \right) \,,
    \label{eq:B}
\end{align}
where $M$ denotes the typical heavy-particle mass scale, namely that of $\psi_{A/B}$ or $S$. 
Explicit expressions for $B_0(m_A,m_B,m_S)$ and $B_1(m_A,m_B,m_S)$ are given in Ref.~\cite{Banno:2026hsc}. 

Let us compare the full result with the contribution induced only through the EW-Weinberg operator in the special case $m_A = m_B = m_S$. 
The full three-loop result is~\cite{Banno:2026hsc}
\begin{align}
    \frac{d_e^{\rm Full}}{e}
    =
    \frac{\alpha_2^2 m_e}{(16 \pi^2)^2}
    \frac{r (r^2-1)}{12}
    \Im (s a^*)
    \left(
    \frac{0.41}{m_A^2}
    +
    \frac{0.22m^2_W}{m_A^4}
    \right)
    +
   \mathcal{O}\left(\dfrac{m_W^4}{m_A^4}\right)\,,
    \label{eq:EDM_Full_same_mass}
\end{align}
where the first term comes from $B_0(m_A,m_B,m_S)$ and the second from $B_1(m_A,m_B,m_S)$ in Eq.~\eqref{eq:eEDM_Full}. 
On the other hand, the contribution mediated only by the EW-Weinberg operator is given by~\cite{Banno:2024apv}
\begin{align}
\label{eq:dEFT_su2}
    \frac{d^{\rm EFT}_e}{e}
    =
    \frac{\alpha_2^2 m_e}{(16 \pi^2)^2}
    \frac{r (r^2-1)}{12}
    \Im (s a^*) 
    \frac{0.14}{m_A^2}\,.
\end{align}
Hence, in the large $m_A$ limit, we obtain $d_{e}^{\rm Full} \simeq 3 d_{e}^{\rm EFT}$. 
This relation remains valid even when the $SU(2)_L$ multiplet masses are not degenerate. 
As a result, the electron EDM constraint in $SU(2)_L$ multiplet models is significantly stronger than the one shown in Fig.~\ref{fig:bau_Lambda}. 
Although Eq.~\eqref{eq:BAULambda_EFT} gives the cutoff range required to explain the BAU within the SMEFT description, the lower bound increases to $57\,{\rm TeV}$ once the full result $d_{e}^{\rm Full} \simeq 3 d_{e}^{\rm EFT}$ is used. 
Thus, the viable BAU window is significantly compressed in the $SU(2)_L$ multiplet models. 

\subsection{Numerical results}

\begin{figure}[t]
    \centering
    \includegraphics[width=0.98\linewidth]{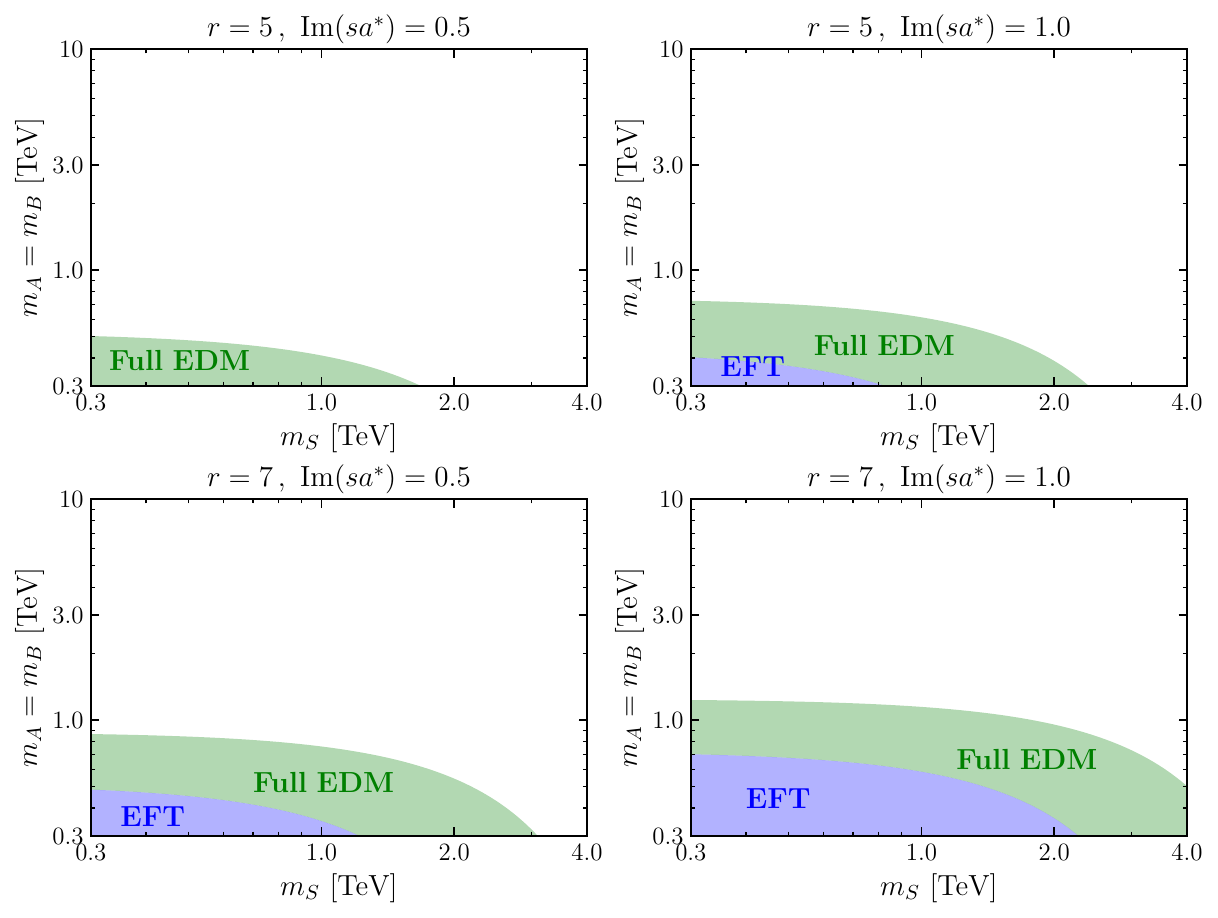}
    \caption{
    Parameter region constrained by electron EDM measurements in the case $m_A=m_B$.
    The blue region is excluded using Eq.~\eqref{eq:EDM_EFT}, while the green region is excluded by the full result in Eq.~\eqref{eq:eEDM_Full}. 
    }
    \label{fig:mAmB_2dim_EFT}
\end{figure}

We now present our numerical results for the production of baryon asymmetry in the $SU(2)_L$ multiplet models. 
In this analysis, we focus on the cases of the EW quintuplet and septuplet fermions with $m_A=m_B$ in the subsequent discussion. 

Figure~\ref{fig:mAmB_2dim_EFT} shows the parameter region probed by the electron EDM measurement for each benchmark point. 
The blue and green regions indicate the parameter space excluded by the current electron EDM constraint using Eq.~\eqref{eq:EDM_EFT} and Eq.~\eqref{eq:eEDM_Full}, respectively. 
To obtain this plot, we used the ancillary file \texttt{solB.txt} provided in Ref.~\cite{Banno:2026hsc}. 
As the figure shows, the full result in Eq.~\eqref{eq:eEDM_Full} is essential to determine the viability of sphalerogenesis in $SU(2)_L$ multiplet models. 

\begin{figure*}[t]
    \centering
    \includegraphics[width=0.98\linewidth]{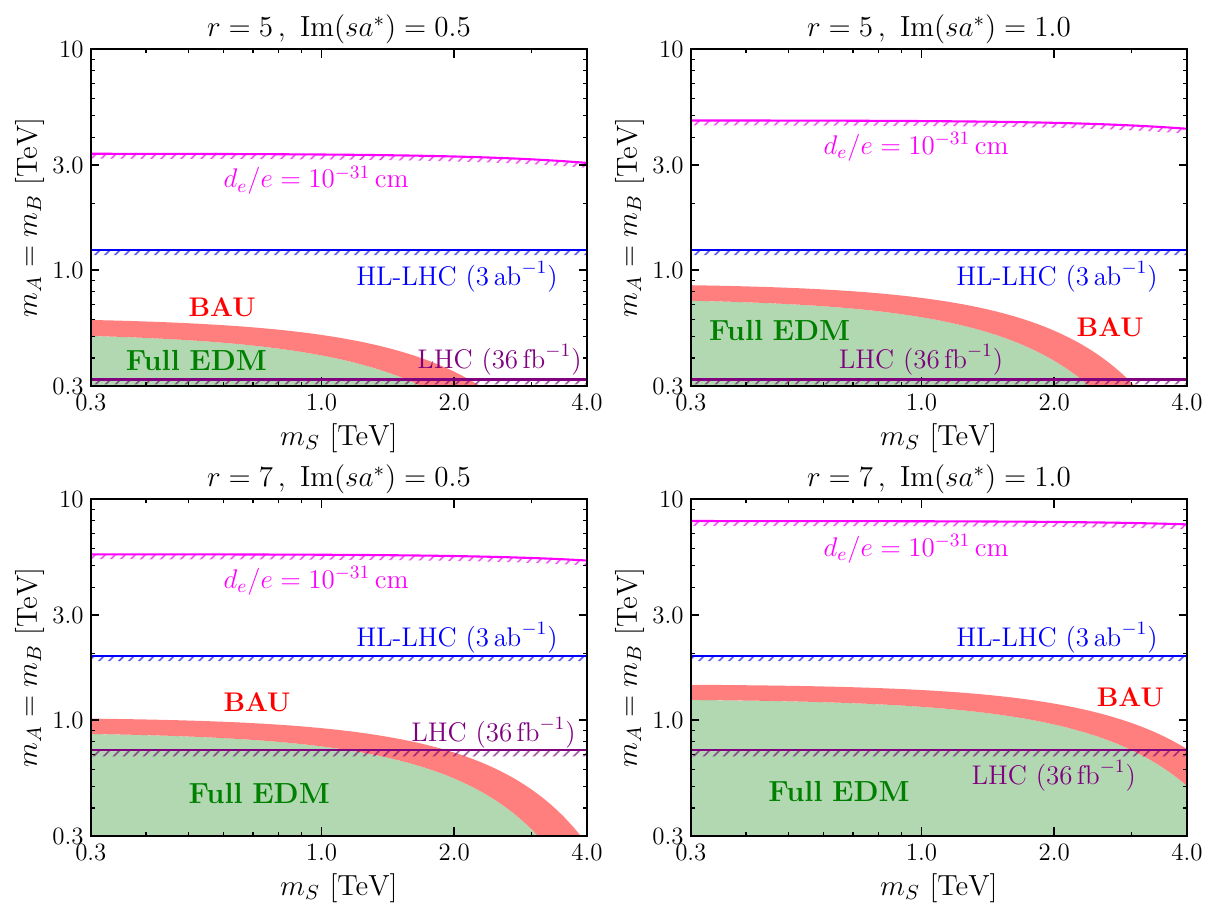}
    \caption{
    Parameter region where the observed BAU can be explained (red region) for each benchmark point within the range for $\kappa_{\rm CP}$ in Eq.~\eqref{eq:kappa_range}. 
    The definition of the green regions is the same as in Fig.~\ref{fig:mAmB_2dim_EFT}. 
    The purple line shows the current indirect bound from mono-lepton searches at the LHC with $\sqrt{s} =13\,{\rm TeV}$ and $36\,{\rm fb}^{-1}$~\cite{Ostdiek:2015aga,Matsumoto:2017vfu,Matsumoto:2018ioi}. 
    The blue line shows the expected bound from the same process at the HL-LHC with $\sqrt{s} =14\,{\rm TeV}$ and $3\,{\rm ab}^{-1}$~\cite{Matsumoto:2018ioi,DiLuzio:2018jwd}. 
    The magenta line corresponds to the contour with $d_{e}/e = 10^{-31}\,{\rm cm}$. 
    }
    \label{fig:summary2}
\end{figure*}

Figure~\ref{fig:summary2} shows the parameter region in which the observed BAU can be explained for each benchmark point. 
In the red region, the measured baryon asymmetry in Eq.~\eqref{eq:BAU} is reproduced within the range in Eq.~\eqref{eq:kappa_range}. 
The green region is excluded by the current electron EDM bound~\cite{Roussy:2022cmp}, as in Fig.~\ref{fig:mAmB_2dim_EFT}. 
The purple and blue lines indicate the current and projected indirect bounds from mono-lepton searches at the LHC and HL-LHC~\cite{Ostdiek:2015aga,Matsumoto:2017vfu,Matsumoto:2018ioi,DiLuzio:2018jwd}. 
Here we show the results for the Majorana-fermion case~\cite{Matsumoto:2018ioi}. 
Fig.~\ref{fig:summary2} indicates that mono-lepton searches at the HL-LHC can thoroughly probe the parameter region in which the BAU is reproduced. 
In addition, if the sensitivity of electron EDM measurements reaches $d_{e}/e = 10^{-31}\,{\rm cm}$, the BAU-favored region will also be thoroughly tested. 
Such sensitivity may be achieved in future experiments such as ACME~III~\cite{Hiramoto:2022fyg}.

Figure~\ref{fig:summary2} shows that our scenario can be tested at the HL-LHC. 
This also suggests that it may be probed in other future collider experiments, such as lepton colliders~\cite{Bottaro:2021srh,Bottaro:2021snn,Kumar:2021umc,Okabe:2023esr,Fukuda:2023yui,Capdevilla:2024bwt} and $100\,{\rm TeV}$ hadron colliders~\cite{Chigusa:2018vxz,DiLuzio:2018jwd,Abe:2019egv}.

\section{Discussions and Conclusions} \label{sec:conclusion}

We have studied the feasibility of sphalerogenesis in the SM extension with new $SU(2)_L$ multiplet fields. 
We have focused on the scenario in which the CP asymmetry in the EW sphaleron process is induced by the EW-Weinberg operator generated from CP-violating Yukawa interactions of the new multiplets. 
Using the reduced description of sphaleron-like processes together with the lattice-based sphaleron-like transition rate, we have estimated the baryon asymmetry produced during the gradual decoupling of CP-violating sphaleron-like processes. 
We then applied this framework to the $SU(2)_L$ multiplet model and identified the parameter regions in which the observed baryon asymmetry can be reproduced.

Our main results are summarized in Fig.~\ref{fig:summary2}.
It shows that viable parameter regions remain consistent with current constraints from the electron EDM measurement and collider searches. 
At the same time, these regions are expected to be thoroughly tested in the near future by improved electron EDM measurements, such as ACME III, as well as by mono-lepton searches at the HL-LHC. 
Therefore, the $SU(2)_L$ multiplet models provide a concrete and phenomenologically testable ultraviolet realization of sphalerogenesis.

Finally, we comment on the implications of our results for DM phenomenology. 
If the $SU(2)_L$ multiplet contains a neutral component, it can be a DM candidate~\cite{Cirelli:2005uq,Cirelli:2007xd,Cirelli:2009uv}. 
When the relic abundance is determined by the conventional freeze-out mechanism, the multiplet fields are typically required to be heavier than $\mathcal{O}(10)\,{\rm TeV}$ for $r \geq 5$~\cite{Mitridate:2017izz,Bottaro:2021snn,Bottaro:2023wjv,Bloch:2024suj}. 
However, Fig.~\ref{fig:summary2} shows that successful sphalerogenesis favors higher $SU(2)_L$ representations, such as $r \geq 5$, with $\mathcal{O}(1)\,{\rm TeV}$ fermions. 
This reveals a clear tension between the BAU-favored region and the conventional freeze-out DM scenario. 
Therefore, explaining both the observed baryon asymmetry and the DM relic abundance within our framework likely requires a non-freeze-out origin of DM. 
We leave such an intriguing direction as a future work.

\section*{Acknowledgement}

The authors thank the Yukawa Institute for Theoretical Physics at Kyoto University, where this work was initiated during the conference YITP-W-25-10 on ``Progress in Particle Physics 2025". 
K.O. is supported by the ``Make New Standards Program for the Next Generation Researchers'' of the Tokai National Higher Education and Research System (THERS).
This work was also financially supported by JST SPRING, Grant Number JPMJSP2125.

\appendix

\section{Sphaleron action \label{sec:sphaleron_action}}

Here, we give the explicit form of the functions $M(\mu)$, $G(\mu)$, and $V(\mu)$ in Eq.~\eqref{eq:Ssph}.
Substituting the sphaleron ansatz in Eq.~\eqref{eq:sph_ansatz} into the action for bosonic fields, the sphaleron action can be expressed by
\begin{align}
\nonumber
S &= \int d^4 x \left[ 
- \frac{1}{2} {\rm tr} \left[ W_{\mu \nu} W^{\mu \nu} \right] + (D_{\mu} \Phi)^{\dagger} D^{\mu} \Phi - V(\Phi) - \frac{g}{3 \Lambda^2} f_{ijk} \widetilde{W}^{i}_{\mu \nu} W^{j\nu \rho} W^{k\mu}_{\rho} \right] \\
&\equiv \int d t \left[ \frac{M(\mu)}{2} \left( \frac{d\mu}{d\eta} \right)^2 + G(\mu) \left( \frac{d\mu}{d \eta} \right)^3 - V(\mu) \right] \,, 
\end{align}
where 
\begin{align}
&\begin{aligned}
M(\mu)
=& ~\frac{4 \pi v(T)}{g}  \int_{0}^{\infty} d \xi \xi^2 
\left[ 
4 \left\{ \frac{4+2c_{\mu}^2}{3} \left( \frac{df}{d\xi} \right)^2 + \frac{4}{\xi^2} \frac{8+2c_{\mu}^2}{3} s_{\mu}^2 f^2 (1-f)^2 \right\} \right. \\ & \left.
+ (1-h)^2 + 2h(1-f)(1-h) + 2 c_{\mu}^2 f(1-h)^2 \right. \\ & \left. 
+\frac{4+2c_{\mu}^2}{3} \left\{ 
h^2 (1-f)^2 + c_{\mu}^2 f(f-2)(1-h)^2 - 2c_{\mu}^2 fh(1-f)(1-h) 
\right\}
\right] \,, 
\end{aligned}
\end{align}
\begin{align}
&\begin{aligned}
G(\mu) 
= \frac{256 \pi}{45} g v(T) s_{\mu}^2 (4 - s_{\mu}^2) \left(\frac{v}{\Lambda}\right)^2 \,, 
\end{aligned} \\
&\begin{aligned}
V(\mu) 
&= \frac{4 \pi v(T)}{g} \int_{0}^{\infty} d\xi \xi^2
\left[ 
\frac{4}{\xi^2} s_{\mu}^2  \left\{ \left( \frac{df}{d\xi}\right)^2 + \frac{2}{\xi^2} f^2 (1-f)^2 s_{\mu}^2 \right\} + \frac{s_{\mu}^2}{2} \left( \frac{dh}{d\xi} \right)^2 \right. \\ & \left. 
\quad + \frac{s_{\mu}^2}{\xi^2} \left\{ h^2 (1-f)^2 - 2 c_{\mu}^2 fh (1-f)(1-h) + c_{\mu}^2 f^2(1-h)^2 \right\} \right. \\ & \left.
\quad + \frac{\lambda}{4g^2}(1-h^2)^2 s_{\mu}^4 \right] \,.
\end{aligned}
\end{align}
We note that the function $G(\mu)$ does not depend on the form of the profile functions $f(\xi)$ and $h(\xi)$. 

\bibliographystyle{JHEP}
\bibliography{reference}

\end{document}